\documentclass[a4,12pt]{article}
\usepackage[dvips]{graphicx,psfrag}
\usepackage{float}
\usepackage{amsmath,amssymb}
\usepackage{multirow}
\usepackage{arydshln}
\usepackage{cite}

\def\ls{\lower0.5ex\hbox{$\buildrel >\over{\scriptstyle\sim}$}}
\def\rs{\lower0.5ex\hbox{$\buildrel <\over{\scriptstyle\sim}$}} 
\allowdisplaybreaks
\begin{document}
\pagestyle{empty} \setlength{\footskip}{2.0cm}
\setlength{\oddsidemargin}{0.5cm}
\setlength{\evensidemargin}{0.5cm}
\renewcommand{\thepage}{-- \arabic{page} --}
\def\mib#1{\mbox{\boldmath $#1$}}
\def\bra#1{\langle #1 |}  \def\ket#1{|#1\rangle}
\def\vev#1{\langle #1\rangle} \def\dps{\displaystyle}
 \def\thebibliography#1{\centerline{REFERENCES}
 \list{[\arabic{enumi}]}{\settowidth\labelwidth{[#1]}\leftmargin
 \labelwidth\advance\leftmargin\labelsep\usecounter{enumi}}
 \def\newblock{\hskip .11em plus .33em minus -.07em}\sloppy
 \clubpenalty4000\widowpenalty4000\sfcode`\.=1000\relax}\let
 \endthebibliography=\endlist
 \def\sec#1{\addtocounter{section}{1}\section*{\hspace*{-0.72cm}
 \normalsize\bf\arabic{section}.$\;$#1}\vspace*{-0.3cm}}
\def\secnon#1{\section*{\hspace*{-0.72cm}
 \normalsize\bf$\;$#1}\vspace*{-0.3cm}}
 \def\subsec#1{\addtocounter{subsection}{1}\subsection*{\hspace*{-0.4cm}
 \normalsize\bf\arabic{section}.\arabic{subsection}.$\;$#1}\vspace*{-0.3cm}}
\vspace*{-1.7cm}
\renewcommand{\labelitemii}{$\circ$}
\begin{flushright}
$\vcenter{
}$
\end{flushright}

\vskip 1.6cm
\begin{center}
  {\Large \bf Correlations in flavor-changing $\mib{tqZ}$ couplings}
\vskip 0.18cm
  {\Large\bf constrained via experimental data}
\end{center}

\vspace{0.5cm}
\begin{center}
\renewcommand{\thefootnote}{\alph{footnote})}
Zenr\=o HIOKI$^{\:1),\:}$\footnote{E-mail address:
\tt hioki@tokushima-u.ac.jp}\ 
Kazumasa OHKUMA$^{\:2),\:}$\footnote{E-mail address:
\tt ohkuma@ice.ous.ac.jp}\ and\
Akira UEJIMA$^{\:2),\:}$\footnote{E-mail address:
\tt uejima@ice.ous.ac.jp}\

\end{center}

\vspace*{0.4cm}
\centerline{\sl $1)$ Institute of Theoretical Physics,\ University of Tokushima}

\centerline{\sl Tokushima 770-8502, Japan}

\vskip 0.2cm
\centerline{\sl $2)$ Department of Information and Computer Engineering,}

\centerline{\sl Okayama University of Science}

\centerline{\sl Okayama 700-0005, Japan}

\vspace*{1.8cm}

\centerline{ABSTRACT}

\vspace*{0.2cm}
\baselineskip=21pt plus 0.1pt minus 0.1pt
Possible non-standard $tqZ$ couplings, where $q=c$ or $u$, originated from general
flavor-changing-neutral-current interactions are studied model-independently using
the effective Lagrangian consisting of several $SU(3)\times SU(2) \times U(1)$
invariant dimension-6 operators. After the electroweak symmetry breaking,
these operators are recombined to form four kinds of independent terms whose
coefficients are complex in general. The Lagrangian could therefore include up to eight independent coupling
parameters. Through searches for experimentally allowed regions of these parameters,
it is found that some correlations exist among the signs and sizes of those couplings.
\vskip 1.5cm

\vfill

\setcounter{page}{0}
\newpage
\renewcommand{\thefootnote}{$\sharp$\arabic{footnote}}
\pagestyle{plain} 
\setcounter{footnote}{0}

Searches for Flavor-Changing Neutral Current (FCNC) 
are a quite attractive mission for future collider experiments:
The existence of physics beyond the standard model is strongly indicated
if new phenomena originated from FCNC are observed, because the event probability
of such phenomena is too tiny to detect within the standard-model framework~\cite{Glashow:1970gm}.
In exploring such rare processes, the top quark is expected to play an especially important role,
since it decays without being affected by non-perturbative effects
thanks to its short lifetime~\cite{Bigi:1980az,Bigi:1986jk} in contrast to the other heavy quarks.

We therefore studied top-quark FCNC processes model-independently in our latest paper \cite{Hioki:2018asl}
and derived constraints on
$tqZ$ ($q=u/c$) couplings, which induce FCNC interactions, using the effective Lagrangian.\footnote{
We have given a detailed list of preceding works by other authors in \cite{Hioki:2018asl}. 
}
We however did not deal with any correlation among those coupling constants there,
so we focus on this issue in this letter and study how these couplings are related with each other,
aiming for deeper understanding of our results.

In our analysis, we use the following effective Lagrangian to describe the general
$tqZ$ interactions~\cite{Buchmuller:1985jz,Arzt:1994gp,AguilarSaavedra:2008zc,Grzadkowski:2010es} :
\begin{alignat}{1}\label{eq:efflag_decay}
  &{\cal L}_{tqZ}  = -\frac{g}{2 \cos \theta_W} 
  \Bigl[\,\bar{\psi}_q(x)\gamma^\mu(f_1^L P_L + f_1^R P_R)\psi_t(x)Z_\mu(x) \Bigr.
  \nonumber\\
 &\phantom{========}
  +\bar{\psi}_q(x)\frac{\sigma^{\mu\nu}}{M_Z}(f_2^L P_L + f_2^R P_R)
   \psi_t(x)\partial_\mu Z_\nu(x) \,\Bigr],
\end{alignat}
where  $g$ and $\theta_W$ are the $SU(2)$ coupling constant and the weak mixing angle,
$P_{L/R}\equiv(1\mp\gamma_5)/2$,
$f_{1/2}^{L/R}$ stand for the non-standard couplings
parameterizing contributions from relevant $SU(3) \times SU(2) \times U(1)$ gauge invariant
dimension-6 effective operators~\cite{Hioki:2018asl,AguilarSaavedra:2008zc}.
We treat these coupling parameters as complex numbers independent of each other
in order to perform analyses as model-independently as possible. 
Thus, the resultant $tqZ$ couplings are expressed by up to eight independent parameters. 

Using the above Lagrangian, we can derive the theoretical partial decay width  
as an eight-variable function ${\mit \Gamma}^{\rm th}_{tqZ}({f_ {1/2}^{L/R}})$.
On the other hand, the experimental partial decay width ${\mit \Gamma}^{\rm exp}_{tqZ}$
is obtained by the product of the branching ratio ${\rm Br}(t \to q Z)$ 
and the top-quark total decay width ${\mit\Gamma}_t$ :
${\mit \Gamma}^{\rm exp}_{tqZ}={\rm Br}(t \to q Z)\times {\mit\Gamma}_t$.
Then, allowed regions for $f_ {1/2}^{L/R}$ are obtained by
varying their real and imaginary parts
at the same time and searching for
the parameter space that satisfies 
$ {\mit \Gamma}^{\rm th}_{tqZ}({f_ {1/2}^{L/R}}) < {\mit \Gamma}^{\rm exp}_{tqZ}$.

Let us briefly show the result for the $tcZ$ couplings as an example (see 
Ref.~\cite{Hioki:2018asl} for more detailed results): 
their allowed regions at 95\% confidence level
for ${\mit\Gamma}_t=1.322$ GeV~\cite{Gao:2012ja}
\footnote{The direct measurement of the total decay width of the top quark 
     is consistent with the prediction by the standard model, but the measured one
     still has a large uncertainty. Therefore, we use the standard-model value here
     instead of the experimental value (see Ref.\cite{Hioki:2018asl} for
     related discussions). Since we are focusing on the rare decays,
     this replacement does not bring any significant problem into our analysis.}
and ${\rm Br}(t \to c Z)<2.3 \times 10^{-4}$~\cite{ATLAS:2017beb}
(thus, we are to solve ${\mit \Gamma}^{\rm th}_{tcZ}<3.0\times 10^{-4}$)
are derived as
 $$
  \left|{\rm Re/Im}(f_{1}^{L/R})\right|\leq 3.4\times 10^{-2},~~
  \left|{\rm Re/Im}(f_{2}^{L/R})\right|\leq 2.8\times 10^{-2}.
$$
Here, we should comment on the meaning of the {\it allowed} regions.
This means that if we give one parameter a value outside its allowed range,
we can no longer reproduce the current experimental data
no matter how we vary the other parameters.

Now, using these results and carrying out some further computations,
we investigate if there is a certain relationship among the couplings.
Let us start with the $t \to cZ$ process.\footnote{The latest upper bound on ${\rm Br}(t \to c Z)$ 
     in the published paper \cite{Aaboud:2018nyl} is not $2.3 \times 10^{-4}$ but a bit weaker
     $2.4 \times 10^{-4}$. We however use the former value because the difference is tiny and
     also because it is crucial for our purpose to provide results that can be directly compared
     with what we gave in \cite{Hioki:2018asl}.}
We first set one of the couplings to its {\it maximum} value and vary all the others.
When ${\rm Re}(f_1^L)$ is taken as such a fixed parameter,
the allowed regions of the remaining couplings
are derived as Table \ref{tab:one_fix_sample1}.
From this table, we can see a relation between ${\rm Re}(f_1^L)$ and ${\rm Re} (f_2^R)$:
the sign of ${\rm Re} (f_2^R)$ is opposite to that of ${\rm Re}(f_1^L)$
and the size of ${\rm Re} (f_2^R)$ is of the same order as ${\rm Re}(f_1^L)$.
Next, in Table \ref{tab:one_fix_sample2}, we show similar bounds to those in
Table \ref{tab:one_fix_sample1}
but in the case that ${\rm Re}(f_2^R)$ is set to its {\it minimum} value.
There we find the same kind of correlation between ${\rm Re}(f_1^L)$ and ${\rm Re}(f_2^R)$ again.

It must be meaningful here to point out the following fact:
While we are now observing the relation between ${\rm Re}(f_1^L)$ and ${\rm Re}(f_2^R)$
assuming that they take the maximal/minimal values, the other parameters can also have
some allowed space though its size is one order of magnitude smaller. In fact, even if
those other parameters had no allowed area, the allowed space for the correlated pair
would not change drastically.

On the other hand, if we assume as an extreme case that only one non-standard coupling exists
in the $tqZ$ interactions, the allowed region of that coupling becomes rather small :
In the case that only ${\rm Re}(f_{1}^{L})$ or ${\rm Re}(f_{2}^{R})$ exists in the $tcZ$ couplings,
we can get the corresponding allowed region as 
$\left|{\rm Re}(f_{1}^{L})\right|\leq 1.9\times 10^{-2}$ or 
$\left|{\rm Re}(f_{2}^{R})\right|\leq 1.5\times 10^{-2}$.

\begin{table}[tb]
\centering
\vspace*{0.4cm}
\caption{Allowed minimum and maximum values of the $tcZ$-coupling parameters for
${\mit\Gamma}^{\rm th}_{tcZ} < 3.0 \times 10^{-4}$
in the case that Re($f_1^L$) is fixed to its maximum value $3.4\times 10^{-2}$.}
\label{tab:one_fix_sample1}
\vspace*{0.3cm}
\begin{tabular}{ccc|cc}
\multicolumn{1}{l}{}     & \multicolumn{2}{c|}{$f_1^L$}                                          & \multicolumn{2}{c}{$f_1^R$}                                      \\ \cline{2-5} 
\multicolumn{1}{l}{}     & Re($f_1^L$)                          & Im($f_1^L)$                    & Re($f_1^R$)                    & Im($f_1^R$)                     \\ \hline
\multicolumn{1}{c}{Min.} & \multirow{2}{*}{$3.4\times 10^{-2}\;$} & $-4.0\times 10^{-3}$           & $-4.0\times 10^{-3}$           & $-4.0\times 10^{-3}$            \\ \cline{1-1}\cline{3-5} 
\multicolumn{1}{c}{Max.} &  {\scriptsize (Fixed)}               & $\phantom{-}4.0\times 10^{-3}$ & $\phantom{-}4.0\times 10^{-3}$ & $\phantom{-}4.0\times 10^{-3}$  \\ \hline
\end{tabular}
\vspace*{0.2cm}
\\
\begin{tabular}{ccc|cc}
\multicolumn{1}{l}{}     & \multicolumn{2}{c|}{$f_2^L$}                                          & \multicolumn{2}{c}{$f_2^R$}                                      \\ \cline{2-5} 
\multicolumn{1}{l}{}     & Re($f_2^L$)                          & Im($f_2^L)$                    & Re($f_2^R$)                    & Im($f_2^R$)                     \\ \hline
\multicolumn{1}{c}{Min.} & $-3.0\times 10^{-3}$                 & $-3.0\times 10^{-3}$           & $-2.5\times 10^{-2}$           & $-3.0\times 10^{-3}$            \\ \hline
\multicolumn{1}{c}{Max.} & $\phantom{-}3.0\times 10^{-3}$       & $\phantom{-}3.0\times 10^{-3}$ & $-2.2\times 10^{-2}$           & $\phantom{-}3.0\times 10^{-3}$  \\ \hline
\end{tabular}
\vspace*{0.4cm}
%
%
\vspace*{0.4cm}
\caption{Allowed minimum and maximum values of the $tcZ$-coupling parameters for
${\mit\Gamma}^{\rm th}_{tcZ} < 3.0 \times 10^{-4}$
in the case that Re($f_2^R$) is fixed to its minimum value $-2.8\times 10^{-2}$.}
\label{tab:one_fix_sample2}
\vspace*{0.3cm}
\begin{tabular}{ccc|cc}
\multicolumn{1}{l}{}     & \multicolumn{2}{c|}{$f_1^L$}                                          & \multicolumn{2}{c}{$f_1^R$}                                      \\ \cline{2-5} 
\multicolumn{1}{l}{}     & Re($f_1^L$)                          & Im($f_1^L)$                    & Re($f_1^R$)                    & Im($f_1^R$)                     \\ \hline
\multicolumn{1}{c}{Min.} & $\phantom{-}2.6\times 10^{-2}$ & $-5.0\times 10^{-3}$           & $-5.0\times 10^{-3}$           & $-5.0\times 10^{-3}$            \\ \hline
\multicolumn{1}{c}{Max.} & $\phantom{-}3.1\times 10^{-2}$       & $\phantom{-}5.0\times 10^{-3}$ & $\phantom{-}5.0\times 10^{-3}$ & $\phantom{-}5.0\times 10^{-3}$  \\ \hline
\end{tabular}
\vspace*{0.2cm}
\\
\begin{tabular}{ccc|cc}
\multicolumn{1}{l}{}     & \multicolumn{2}{c|}{$f_2^L$}                                          & \multicolumn{2}{c}{$f_2^R$}                                      \\ \cline{2-5} 
\multicolumn{1}{l}{}     & Re($f_2^L$)                          & Im($f_2^L)$                    & Re($f_2^R$)                    & Im($f_2^R$)                     \\ \hline
\multicolumn{1}{c}{Min.} & $-4.0\times 10^{-3}$                 & $-4.0\times 10^{-3}$           & \multirow{2}{*}{$\,\;-2.8\times 10^{-2}\;$} & $-4.0\times 10^{-3}$            \\ \cline{1-3}\cline{5-5}
\multicolumn{1}{c}{Max.} & $\phantom{-}4.0\times 10^{-3}$       & $\phantom{-}4.0\times 10^{-3}$ &  {\scriptsize (Fixed)} & $\phantom{-}4.0\times 10^{-3}$  \\ \hline
\end{tabular}
\vspace*{0.8cm}
\end{table}

We performed these analyses for all the remaining parameters, too.
The results are summarized as follows:
\begin{itemize}
\item Relational expressions ${\rm Re/Im}(f_{1/2}^{L/R})=-{C\,\rm Re/Im}(f_{2/1}^{R/L})$ hold,
where the maximum ranges of $f_1^{L/R}$ and $f_2^{L/R} $ are given
with $0.93 \lesssim C \lesssim 1.1$ and $0.65 \lesssim C \lesssim 0.73$ respectively by
substituting the maximum or minimum values of
$f_2^{R/L}$ and $f_1^{R/L}$ in the right-hand side of the expressions.
\item Three or more non-standard couplings cannot take large values (within the allowed regions) at the same time. 
\item The allowed region could be roughly twice larger when two non-standard coupling
constants exist than in the case that only one non-standard coupling constant exists.
\end{itemize}

\begin{table}[H]
	\centering
	\vspace*{0.4cm}
	\caption{Allowed minimum and maximum values of the $tuZ$-coupling parameters for ${\mit\Gamma}^{\rm th}_{tuZ} < 2.2 \times 10^{-4}$
		in the case that Re($f_1^L$) is fixed to its maximum value $2.9\times 10^{-2}$.}
	\label{tab:one_fix_sample3}
	\vspace*{0.3cm}
	\begin{tabular}{ccc|cc}
		\multicolumn{1}{l}{}     & \multicolumn{2}{c|}{$f_1^L$}                                          & \multicolumn{2}{c}{$f_1^R$}                                      \\ \cline{2-5} 
		\multicolumn{1}{l}{}     & Re($f_1^L$)                          & Im($f_1^L)$                    & Re($f_1^R$)                    & Im($f_1^R$)                     \\ \hline
		\multicolumn{1}{c}{Min.} & \multirow{2}{*}{$2.9\times 10^{-2}\;$} & $-4.0\times 10^{-3}$           & $-4.0\times 10^{-3}$           & $-4.0\times 10^{-3}$            \\ \cline{1-1}\cline{3-5} 
		\multicolumn{1}{c}{Max.} &  {\scriptsize (Fixed)}               & $\phantom{-}4.0\times 10^{-3}$ & $\phantom{-}4.0\times 10^{-3}$ & $\phantom{-}4.0\times 10^{-3}$  \\ \hline
	\end{tabular}
	\vspace*{0.2cm}
	\\
	\begin{tabular}{ccc|cc}
		\multicolumn{1}{l}{}     & \multicolumn{2}{c|}{$f_2^L$}                                          & \multicolumn{2}{c}{$f_2^R$}                                      \\ \cline{2-5} 
		\multicolumn{1}{l}{}     & Re($f_2^L$)                          & Im($f_2^L)$                    & Re($f_2^R$)                    & Im($f_2^R$)                     \\ \hline
		\multicolumn{1}{c}{Min.} & $-4.0\times 10^{-3}$                 & $-4.0\times 10^{-3}$           & $-2.2\times 10^{-2}$           & $-4.0\times 10^{-3}$            \\ \hline
		\multicolumn{1}{c}{Max.} & $\phantom{-}4.0\times 10^{-3}$       & $\phantom{-}4.0\times 10^{-3}$ & $-1.8\times 10^{-2}$           & $\phantom{-}4.0\times 10^{-3}$  \\ \hline
	\end{tabular}
	\vspace*{0.4cm}
	%
	%
	\vspace*{0.4cm}
	\caption{Allowed minimum and maximum values of the $tuZ$-coupling parameters for ${\mit\Gamma}^{\rm th}_{tuZ} < 2.2 \times 10^{-4}$
		in the case that Re($f_2^R$) is fixed to its minimum value $-2.4\times 10^{-2}$.}
	\label{tab:one_fix_sample4}
	\vspace*{0.3cm}
	\begin{tabular}{ccc|cc}
		\multicolumn{1}{l}{}     & \multicolumn{2}{c|}{$f_1^L$}                                          & \multicolumn{2}{c}{$f_1^R$}                                      \\ \cline{2-5} 
		\multicolumn{1}{l}{}     & Re($f_1^L$)                          & Im($f_1^L)$                    & Re($f_1^R$)                    & Im($f_1^R$)                     \\ \hline
		\multicolumn{1}{c}{Min.} & $\phantom{-}2.2\times 10^{-2}$       & $-4.0\times 10^{-3}$           & $-4.0\times 10^{-3}$           & $-4.0\times 10^{-3}$            \\ \hline
		\multicolumn{1}{c}{Max.} & $\phantom{-}2.7\times 10^{-2}$       & $\phantom{-}4.0\times 10^{-3}$ & $\phantom{-}4.0\times 10^{-3}$ & $\phantom{-}4.0\times 10^{-3}$  \\ \hline
	\end{tabular}
	\vspace*{0.2cm}
	\\
	\begin{tabular}{ccc|cc}
		\multicolumn{1}{l}{}     & \multicolumn{2}{c|}{$f_2^L$}                                          & \multicolumn{2}{c}{$f_2^R$}                                      \\ \cline{2-5} 
		\multicolumn{1}{l}{}     & Re($f_2^L$)                          & Im($f_2^L)$                    & Re($f_2^R$)                    & Im($f_2^R$)                     \\ \hline
		\multicolumn{1}{c}{Min.} & $-4.0\times 10^{-3}$                 & $-4.0\times 10^{-3}$           & \multirow{2}{*}{$\,\;-2.4\times 10^{-2}\;$} & $-4.0\times 10^{-3}$            \\ \cline{1-3}\cline{5-5}
		\multicolumn{1}{c}{Max.} & $\phantom{-}4.0\times 10^{-3}$       & $\phantom{-}4.0\times 10^{-3}$ &  {\scriptsize (Fixed)} & $\phantom{-}4.0\times 10^{-3}$  \\ \hline
	\end{tabular}
	\vspace*{0.8cm}
\end{table}

In the same way, we then analyzed the $t \to u Z$ process. The main results 
coming from the current upper bound
${\rm Br}(t \to uZ)<1.7 \times 10^{-4}$~\cite{ATLAS:2017beb,Aaboud:2018nyl} are shown in
Table \ref{tab:one_fix_sample3} and Table \ref{tab:one_fix_sample4},
which are to be compared with Table \ref{tab:one_fix_sample1} and Table \ref{tab:one_fix_sample2}.
There we can clearly see parameter-correlations quite similar to those in the $tcZ$ couplings.
That is, what we found here is common to both the $tcZ$ couplings and the $tuZ$ couplings.

Finally, we would like to present important comments on (1) to what extent our results
depend on the upper limit of the branching ratio of $t \to q Z$ and
(2) how model-independent and unique our analyses are:\\
(1) We used here ${\rm Br}(t \to q Z)$ presented in \cite{ATLAS:2017beb,Aaboud:2018nyl},
but it is expected to be improved by half
at High-Luminosity Large Hadron Collider~\cite{CMS-PAS-FTR-13-016}.
Therefore we performed the same computations assuming that reduced upper bound.
Of course, the allowed parameter ranges are thereby narrowed, which we already showed
in the previous work~\cite{Hioki:2018asl}, but interestingly enough we found that
the coupling-parameter correlations and related results
are little affected and still hold even there.\\
(2) We worked in the framework of the effective Lagrangian consisting of dimension-6 operators
and treated all their coefficients as independent complex numbers, which led to eight-parameter analyses.
That is, our results were derived by varying those eight parameters at the same time and comparing the
resultant ${\mit\Gamma}^{\rm th}_{tqZ}$ with the corresponding experimental bound.
This is quite in contrast with preceding many analyses, where only some coefficients are treated as free
parameters at once fixing the others. For example, the CMS Collaboration carried out detailed studies in
\cite{Sirunyan:2017kkr} but they dropped some terms in the effective Lagrangian from the beginning.
The ATLAS Collaboration also performed comprehensive analyses in \cite{Aaboud:2018nyl}, but what they gave
are constraints on the absolute values of four coefficients assuming that only one parameter has a non-zero
value in each analysis. In such limited analyses, we cannot take into account possible cancellations
among contributions from different parameters. Indeed the existence of the correlations shown here tells us
that it is true.
%


In conclusion, we have studied possible non-standard $tqZ$ couplings and correlations
among them in the framework of the effective Lagrangian.
This Lagrangian can incorporate up to eight independent coupling parameters
describing FCNC interactions.
The allowed regions of these couplings satisfying
the current experimental limits get larger when several numbers of
couplings exist than when only one coupling exists.
It was found that the allowed region becomes the largest when there are relations as
${\rm Re/Im}(f_{1/2}^{L/R})=-C\,{\rm Re/Im}(f_{2/1}^{R/L})$ with
$C \simeq 0.65 \sim 0.73$ ($C \simeq 0.93 \sim 1.1$) in the case that the maximum or minimum value
of $f_1^{R/L}$ ($f_2^{R/L}$) is substituted in the right-hand side.
Tables \ref{tab:one_fix_sample1} -- \ref{tab:one_fix_sample4} also tell us that
the non-standard couplings except for the two correlated ones
can take non-zero values as well though their sizes are one order of magnitude smaller
than those of the correlated ones.
However, even if they had no allowed regions,
the current experimental limits could be realized by the existence of two correlated
couplings alone.

Our results are from the current experimental data, but it is quite interesting to
note that the parameter correlations and related results are little affected even if
the experimental limits on ${\rm Br}(t \to q Z)$ are improved, e.g., by half at future facilities.
Since this analysis was performed in a very general framework and does not depend
on any special assumptions,
the results pointed out here will be useful information
for constructing a specific model inducing rather strong FCNC interactions.
%
\secnon{Acknowledgments}
%
This work was partly supported by the Grant-in-Aid for Scientific Research (C) 
Grant Number 17K05426 from the Japan Society for the Promotion of Science.


\baselineskip=20pt plus 0.1pt minus 0.1pt

\vspace*{0.8cm}

\end{document}